\documentclass[12pt,preprint]{aastex}

\slugcomment{appeared in ApJ, 636:261--266}

\shorttitle{The Dense Molecular Cloud G0.11-0.11}
\shortauthors{Handa et al.}

\begin{document}


\title{
THERMAL SiO AND H$^{13}$CO$^+$ LINE OBSERVATIONS OF
THE DENSE MOLECULAR CLOUD G0.11$-$0.11 IN THE GALACTIC CENTER REGION
\thanks{
This work was carried out under the common-use observation program at 
Nobeyama Radio Observatory (NRO).
}
}


\author{T. Handa}
\affil{Institute of Astronomy, University of Tokyo,
    Osawa 2-21-1, Mitaka, Tokyo 181-0011, Japan}
\email{handa@ioa.s.u-tokyo.ac.jp}

\author{M. Sakano}
\affil{Department of Physics and Astronomy, University of Leicester, 
     Leicester LE1 7RH, UK}
\email{mas@star.le.ac.uk}

\author{S. Naito, M. Hiramatsu}
\affil{Department of Astronomy, University of Tokyo, 
     Hongo 7-3-1, Bunkyo, Tokyo 113-0033, Japan}
\email{snaito@ioa.s.u-tokyo.ac.jp, hiramats@alma.mtk.nao.ac.jp}

\and

\author{M. Tsuboi}
\affil{Nobeyama Radio Observatory, National Astronomical Observatory Japan,
Nobeyama, Minami-Saku, Nagano 384-1305, Japan;
Department of Astronomical Science, Graduate University for Advanced Studies
(Soken-Dai), Mitaka, Tokyo 181-8588, Japan;
and Department of Astronomy, University of Tokyo,
Hongo 7-3-1, Bunkyo-ku, Tokyo 113-0033, Japan}
\email{tsuboi@nro.nao.ac.jp}




\begin{abstract}
We obtained the first view in H$^{13}$CO$^+$ $J=1-0$ and 
a high-resolution map in thermal SiO lines of G0.11$-$0.11,
which is a molecular cloud situated between 
the Galactic Center radio arc and Sgr A.
From a comparison with previous line observations, we found
that the H$^{13}$CO$^+$ $J=1-0$ line is optically thin,
whereas the thermal SiO lines are optically thick.
The line intensity in H$^{13}$CO$^+$ $J=1-0$ shows that
the cloud has a large column density, up to 
$N(\mathrm{H}_2)=(6-7)\times10^{23} \mathrm{cm^{-2}}$,
which corresponds to about 640--740 mag in $A_{\mathrm{V}}$
or 10--12 mag in $A_{25\mu\mathrm{m}}$.
The estimated column density is the largest known of any 
even in the Galactic center region.
We conclude from the intensity ratio of SiO $J=1-0$ to CS $J=1-0$ that
emitting gas is highly inhomogeneous for SiO abundance on 
a scale smaller than the beam width $\sim$35\arcsec.
\end{abstract}





\section{INTRODUCTION}

Recent infrared observations have revealed the existence of a
population of dense bright-star clusters
in the Galactic center region (e.g., \citet{figer2002}),
such as the Arches cluster or the Quintuplet cluster.
Although these clusters must be made from molecular clouds
in the Galactic center region,
it remains unclear
what modifies the star formation process to form such extremely massive stars.
Further detailed observations of dense molecular clouds
in the Galactic center region
may provide a key to understanding the physical properties of the clouds and their relationship to this star-forming mechanism.

Large-scale surveys in molecular lines have revealed that
molecular clouds in the central $\sim$~100 pc of the Galaxy are
different from those in the Galactic disk.
For example, the incidence of relatively dense clouds is higher in the Galactic center region.
The Nobeyama Radio Observatory (NRO) CS survey has shown that density of
most molecular clouds there
is over $10^4~\mathrm{cm}^{-3}$ \citep{tsuboi1999}. 
Moreover, the multiline observations show that 
even the CO-emitting region (or outer envelope) of a typical Galactic center molecular cloud 
is under high pressure at 
$n T_\mathrm{K} = 10^5~\mathrm{K~cm^{-3}}$ 
\citep{oka1998,sawada2001}.

Among dense molecular clouds in the Galactic center region, 
the molecular cloud G0.11$-$0.11\footnote{
The cloud is referred to as Tsuboi's shell in the NRO 45-m CO survey 
\citep{oka2001}, as G0.11$-$0.11 in \citet{reich2003},
and as the TUH shell in \citet{yusefzadeh2002}.} 
is unique and one of the most interesting objects.
It is located between Sgr~A and the Galactic center arc (GCA).
The NRO CS survey \citep{tsuboi1999} shows that 
G0.11$-$0.11 is well separated in $l$-$b$-$v$ space from 
the main ridge of the CS emission,
which is the central disk of molecular gas in the Galactic center.
The CS observations reveal 
that G0.11$-$0.11 has a large molecular mass and large velocity width 
at the eastern\footnote{
In this paper, all directions on the sky are in terms of Galactic coordinates.}
 and western edges of the cloud.
The eastern edge appears to have an interaction with the GCA \citep{tsuboi1997}. 
G0.11$-$0.11 is bright at the X-ray fluorescent iron line \citep{yusefzadeh2002},
which suggests that dense gas in the cloud reflects X-rays
from an intense source.

The morphologies of G0.11$-$0.11 in CS and CO lines are similar.  
These lines are presumably optically thick at the surface of the cloud. 
To obtain the physical properties of the cloud, 
observations in an optically thin line are required.
The H$^{13}$CO$^+$ ($J=1-0$) line should be suitable
because of its very low abundance.
Here we present the first view in an H$^{13}$CO$^+$ line 
of G0.11$-$0.11,
exploring the molecular gas distribution in the cloud.
At the same time, we present a high-resolution view in 
thermal SiO lines.
The thermal SiO lines are thought to be a good tracer of hot and 
shocked regions
because this molecule is in gas phase only under high-temperature environment
\citep{ziurys1989}. 
For example, thermal SiO emission is detected in bipolar flow sources 
and in shocked shells of supernova remnants. 
\citet{martinpintado1997} surveyed the Galactic center region
in the thermal SiO line.
However, they observed only half the extent of G0.11$-$0.11, and 
with poor angular resolution.
We have now made observations of G0.11$-$0.11 
in these lines using the Nobeyama 45 m telescope
with much higher resolution.

\section{OBSERVATIONS}


   We have observed G0.11$-$0.11 in April 2002,
using the Nobeyama 45 m telescope, 
simultaneously observing at the spectral lines of 
H$^{13}$CO$^+$ $J=1-0$ (86.754330 GHz),
SiO $J=1-0, v=0$ (43.423798 GHz), and SiO $J=2-1, v=0$ (86.846998 GHz).  
The FWHM beam sizes at 43 and 86 GHz are 35\arcsec~ and 18\arcsec, 
respectively.
The receiver front ends were SIS receivers at 43 and 86 GHz
with a polarization splitter.
The observed region is a rectangular area of
$0\degr~4\arcmin ~\le ~l ~\le ~0\degr10\arcmin$, and $-0\degr10\arcmin ~\le
~b ~\le~ -0\degr4\arcmin$,
which covers the whole cloud.
The spacing of the observation grid is 20\arcsec,
which corresponds to 0.82 pc
at the distance to the Galactic center, 8.5 kpc.
The main-beam efficiencies at 43 and 86 GHz are
0.81 and 0.50, respectively.
Two SiO maser sources, OH~2.6$-$0.4 and VX Sgr, were observed
every hour in order to check the telescope pointing.
Typical pointing accuracy was 5\arcsec~ during this observation.

   We used acousto-optic spectrometers with 250 MHz bandwidth, of
which the respective velocity resolutions at 43 and 86 GHz
are 0.87 $\mathrm{km~s^{-1}}$ and 0.44 $\mathrm{km~s^{-1}}$, respectively.
The line intensities were calibrated by the chopper wheel method
\citep{kutner1981} 
in order to correct the antenna temperature for atmospheric attenuation,
$T_{\mathrm{A}}^*$.
The system temperatures during this observation were 
300 K at 43 GHz and 500 K at 86 GHz.
Linear or parabolic baselines were applied to all the spectra.




\section{RESULTS}
\subsection{Features and Morphology of the Cloud}

\begin{figure}
\epsscale{.80}
\plotone{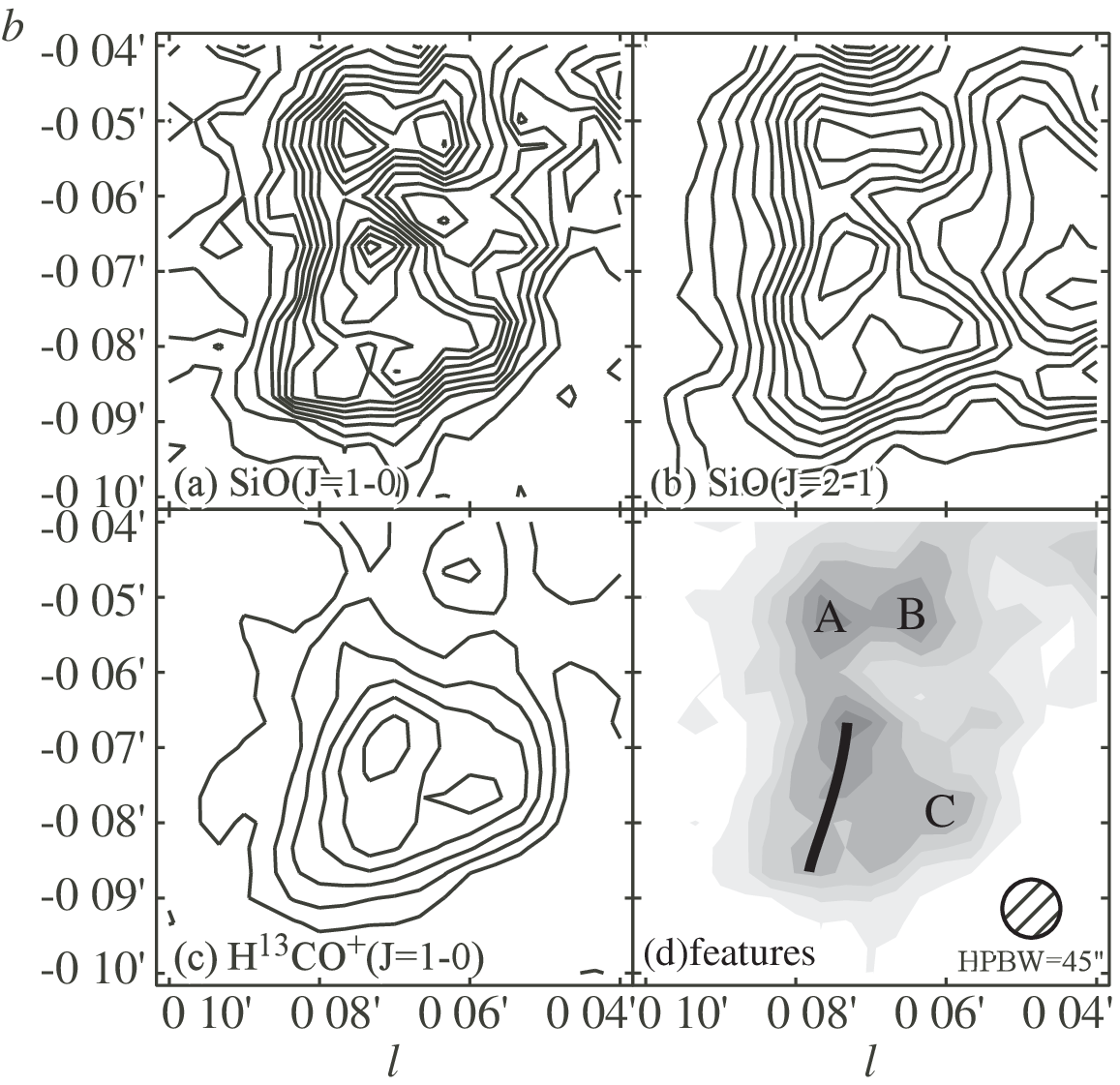}
\caption{
Integrated intensity maps of G0.11$-$0.11 in (a) SiO $J=1-0, v=0$, 
(b) SiO $J=2-1, v=0$, and (c) H$^{13}$CO$^+$ $J=1-0$,
and (d) a schematic chart of main features.  
The velocity range is $15 ~\le ~V_{\mathrm{LSR}} ~\le ~45 ~\mathrm{km~s^{-1}}$.
The FWHM beam sizes are adjusted to 45\arcsec, shown in (d).
The intensity scale is in integrated main-beam brightness temperature, 
$\int ~T_{\mathrm{MB}} ~dv$. 
Both the first contour level and the contour interval are 
2 K $\mathrm{km~s^{-1}}$ for (a--c).
The gray scale in (d) is the same map as in (a).
}\label{FigIntMap}
\end{figure}

  Figure \ref{FigIntMap} shows the integrated intensity maps 
in the three lines in the velocity range of
$15~\mathrm{km~s^{-1}} ~\le ~v_{\mathrm{LSR}} ~\le ~45~\mathrm{km~s^{-1}}$.
The spatial resolutions are adjusted to 
45\arcsec~ by applying Gaussian convolution.
A shell-like molecular cloud is seen in all the three lines.
The appearance in the SiO lines resembles that in the CS $J=1-0$ line
(see Fig.~1 of Tsuboi et al. 1997).

However, several differences are apparent in the images 
in the H$^{13}$CO$^+$ $J=1-0$ and SiO lines.
The H$^{13}$CO$^+$ $J=1-0$ intensity is significantly concentrated
to the southern half of the cloud,
although the cloud seems to extend beyond $b~\geq~-0\degr6\arcmin$
in SiO and CS images.
In the H$^{13}$CO$^+$ $J=1-0$ line,
the integrated intensity in $b~\leq~-0\degr6\arcmin$ is 80 \% of 
the total intensity of the whole cloud, which is integrated over
$0\degr5\arcmin~\leq~l~\leq~0\degr9\arcmin$,
$-0\degr9\arcmin20\arcsec~\leq~b~\leq~-0\degr4\arcmin20\arcsec$.
These discrepancies are presumably 
due to the difference in optical depths 
between the SiO and H$^{13}$CO$^+$ lines 
(see detail in \S~\ref{sec:nh}).
Namely, the H$^{13}$CO$^+$ $J=1-0$ line intensity traces 
the column density, but the SiO lines do not.
Thus, G0.11$-$0.11 shows significant difference in column density
below and above a front at $b~=~-0\degr6\arcmin$.

G0.11$-$0.11 shows four distinctive features in these lines.
Along the eastern edge of the cloud, 
a prominent ridge is seen in all the three lines.
We dub it the E ridge hereafter 
(Fig. ~\ref{FigIntMap}, \textit{solid line}).
At $l=0\degr7\arcmin$, 
it extends from $b=-0\degr6\arcmin$ to $b=-0\degr9\arcmin$ 
perpendicular to the Galactic plane. 
On the northern end of the E ridge, a peak is seen in both the SiO lines
at $l=0\degr 8\arcmin, b=-0\degr 5\arcmin 20\arcsec$~ (hereafter peak A).
At $l=0\degr 6\arcmin 20\arcsec, b=-0\degr 5\arcmin 20\arcsec$,
another peak (peak B) is seen in the SiO lines.
The other prominent feature is a peak 
at $l=0\degr 6\arcmin, b=-0\degr 8\arcmin$ (peak C). 
Figure~\ref{FigIntMap} illustrates these features.

\begin{figure}
\plotone{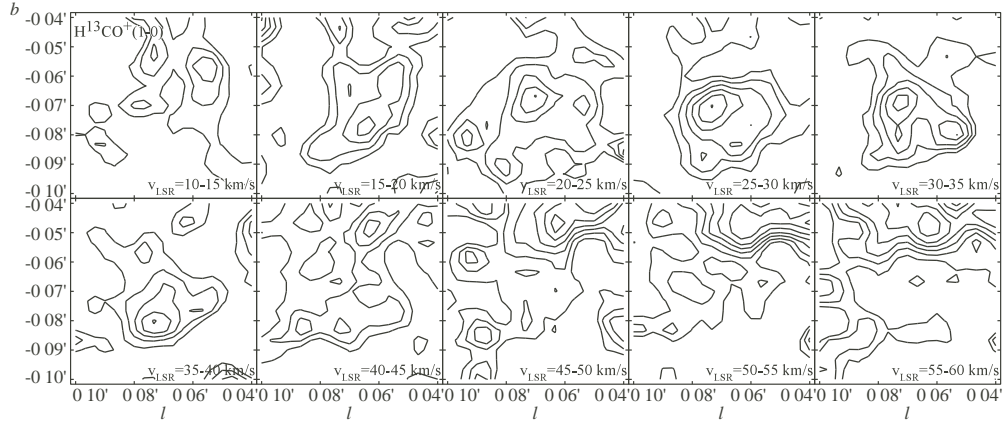}
\caption{
Channel map of intensity in H$^{13}$CO$^+$ $J=1-0$.
The data are smoothed by Gaussian to 45\arcsec~ resolution.
The intensity scale is in main-beam brightness temperature, $T_{\mathrm{MB}}$. 
Both the first contour level and contour interval are 0.1 K
in all panels.
}\label{FigH13CO10channelMap}
\end{figure}

Figures~\ref{FigH13CO10channelMap}, ~\ref{FigSiOchannelMap}, and
~\ref{FigSiO21channelMap} show the channel maps
of the H$^{13}$CO$^+$ $J=1-0$ line and
the two SiO lines with a velocity interval of 5 $\mathrm{km~s^{-1}}$.
The FWHM in these figures is increased to be 45\arcsec~ by Gaussian smoothing.
Typical rms noise levels in $T_{\mathrm{MB}}$ for
H$^{13}$CO$^+$ $J=1-0$, SiO $J=1-0$, and $J=2-1$ lines are 0.042 K,
0.050 K, and 0.042 K, respectively.

The E ridge is seen in the SiO channel maps between
$15~\mathrm{km~s^{-1}}~<~v_{\mathrm{LSR}}~<~50 ~\mathrm{km~s^{-1}}$.
The E ridge extends for 3\arcmin~ in Galactic latitude,
equivalent to 7 pc at a distance of 8.5~kpc.
A corresponding feature is also seen in the CS line 
(Fig.~2 in Tsuboi et al. 1997).
The E ridge is extended in the direction parallel to the GCA.
This morphology may suggest an interaction of the molecular gas with the GCA.
However, this interaction is probably not very strong, if it exists,
because the E ridge is not the most prominent feature in the SiO image.

In the H$^{13}$CO$^+$ $J=1-0$ map the E ridge is also distinguishable,
but less prominent than in the SiO lines,
and very weak in $b ~\ge ~-0\degr6\arcmin$.
The velocity structures of the E ridge 
in the SiO and H$^{13}$CO$^+$ lines are similar at
$v_{\mathrm{LSR}}~\le~40~\mathrm{km~s^{-1}}$.
The SiO emission is extended beyond 
$v_{\mathrm{LSR}}~\ge~40~\mathrm{km~s^{-1}}$,
but the H$^{13}$CO$^+$ $J=1-0$ emission is not.
We note that this extension of the E ridge in the SiO emission connects
at $v~>~45 ~\mathrm{km~s^{-1}}$ to the ridge extending at
$b=-0\degr 5\arcmin$, which goes through peak B
(see Fig.~\ref{FigH13CO10_lv}).

Peak A is seen in the SiO maps between
$10~\mathrm{km~s^{-1}}~<~v_{\mathrm{LSR}}~<~35 ~\mathrm{km~s^{-1}}$.
At the high-redshift end, 
peak A is merged into the ridge 
along the Galactic plane through peak B.
In the SiO line images, 
peak A appears to be somehow connected with the E ridge.
However, the H$^{13}$CO$^+$ line image shows
no feature corresponding to peak A or B
($v_{\mathrm{LSR}}~<~35~\mathrm{km~s^{-1}}$), whereas
it seems to show the E ridge.
Hence, peak A is unlikely to be a part of the E ridge.

Peak B is seen in the SiO maps at
$v_{\mathrm{LSR}}~>~30 ~\mathrm{km~s^{-1}}$.
Beyond $v_{\mathrm{LSR}}~>~45 ~\mathrm{km~s^{-1}}$
the position of peak B is shifted northward by 20\arcsec.
In the H$^{13}$CO$^+$ map a clear counterpart is seen
only beyond $v_{\mathrm{LSR}}~>~45 ~\mathrm{km~s^{-1}}$.
This suggests that peak B may be a double source and separable
at $v_{\mathrm{LSR}}~=~45 ~\mathrm{km~s^{-1}}$.
In any case, peak B with $v_{\mathrm{LSR}}~<~45 ~\mathrm{km~s^{-1}}$
is only seen in the SiO lines.

Peak C is seen in the range
$30~\mathrm{km~s^{-1}}~<~v_{\mathrm{LSR}}~<~45 ~\mathrm{km~s^{-1}}$.
The H$^{13}$CO$^+$ map shows its counterpart clearly.
Peak C is morphologically connected to the E ridge in $l-b-v$ space.
The E ridge and peak C might be two main parts of G0.11$-$0.11.

Another prominent ridge with $v~>~45 ~\mathrm{km~s^{-1}}$ at
$b=-0\degr 5\arcmin$ parallel to the Galactic plane 
is seen in all the observed lines.
The large-scale velocity structure observed
in CS \citep{tsuboi1999} suggests that
this ridge is a blueshifted wing of the main ridge
of the Galaxy through the whole Galactic center region.
Therefore, we do not discuss this feature in this paper.

\subsection{Intensity Ratio \label{sec:ratio}}
To evaluate the morphological resemblance among the SiO lines and 
difference between the H$^{13}$CO$^+$ and SiO lines quantitatively,
we estimate intensity ratios of observed lines, which are keys to determine
the optical depth and/or physical conditions of the emitting gas
in G0.11$-$0.11.
First, we estimate an intensity ratio of two SiO lines,
$R_{\mathrm{SiO(2-1)/SiO(1-0)}}$.
To calculate an average value, 
we use an intensity correlation for all the observed points
in a box assigned in $l-b-v$ space for each feature.
To remove the difference in resolution 
due to different beam size at the three lines,
we reduce the resolution to be 45\arcsec~ by appropriate Gaussian convolution.
We found the ratios of the SiO $J=2-1$ line to the SiO $J=1-0$ line to be 
0.9--1.0 for the E ridge and the three peaks and also found 
no significant difference in the ratios among the regions in the cloud.

We also estimated the line intensity ratios of 
H$^{13}$CO$^+$ $J=1-0$ to SiO $J=1-0$
and found them to be uniform in each feature,
although they differ significantly
between the northern and southern parts of G0.11$-$0.11. 
For the E ridge and peak C, they are 0.5.
For peaks A and B, they are about 0.2 or smaller,
although the signal-to-noise ratio is poor.

\section{DISCUSSION}
\subsection{Column Density and Mass of the Cloud \label{sec:nh}}
The H$^{13}$CO$^+$ $J=1-0$ line is expected to be optically thin, 
because of its small abundance.
We can check it by comparing the H$^{13}$CO$^+$ $J=1-0$ intensity
with the CS $J=2-1$ intensity.
We should note that
since the excitation parameters of both the lines are similar, 
their intensity ratio ought not to be a strong function of 
the physical conditions of the gas; 
accordingly, the only causes of variation in this ratio 
must be variations either in the relative abundances of 
the species or in their relative optical depth.
Using an H$^{13}$CO$^+$ abundance of $10^{-10}$
and a CS abundance of $10^{-8}$
\citep{garciaburillo2000,irvine1987}
together with excitation parameters of the lines,
the expected intensity ratio of H$^{13}$CO$^+$ $J=1-0$ to
CS $J=2-1$ is about $6 \times 10^{-3}$,
if both lines are optically thin.
Using CS observations by \citet{tsuboi1997},
the line intensity ratio of H$^{13}$CO$^+$ $J=1-0$
to CS $J=2-1$ is calculated to be 0.12--0.14 
in the southern part of G0.11$-$0.11.
It follows that the CS line in this locality must be optically thick, 
whereas the optical depth of H$^{13}$CO$^+$ $J=1-0$ is
about 0.1.
In the northern part of the cloud, 
the ratio is about 0.05 or smaller
and the H$^{13}$CO$^+$ $J=1-0$ line is therefore optically thin.

Then we estimate the column density of
the southern part of G0.11$-$0.11
and the molecular mass of the whole cloud
from H$^{13}$CO$^+$ $J=1-0$ intensity
under the condition of local thermal equilibrium (LTE).
In this case, 
we need to know the kinetic temperature ($T_{\mathrm{K}}$) of the emitting gas. 
The kinetic temperature of molecular gas in the Galactic
center region is controversial.  
In the Galactic center region, 
$T_{\mathrm{K}}$ of dense molecular clouds is estimated to be 
60--80 K or hotter \citep{morris1983,huettemeister1993,lis2001}.
Here we assume $T_{\mathrm{K}}=70~\mathrm{K}$.
Thus, the column density of molecular hydrogen at the E ridge is
derived to be $N(\mathrm{H}_2)=(6-7)\times10^{23} \mathrm{cm^{-2}}$.  
This corresponds to about 640--740 mag in $A_{\mathrm{V}}$
(visual extinction) at the typical gas-to-dust ratio 
expected for dense clouds, and about 80--90 mag and 10--12 mag
in $A_{\mathrm{K}}$ and $A_{\mathrm{25\mu m}}$
(extinction at $25\mu m$), respectively
\citep{mathis2000}.
Similar values are obtained for peak C.
This large $A_{\mathrm{25\mu m}}$ is consistent with the fact that 
G0.11$-$0.11 is observed as a shadow in an infrared map 
with the \textit{Midcourse Space Experiment} 
(\textit{MSX}; Egan et al. 1998). 
The shadow has a spatial extension on the sky
similar to that in $\mathrm{H^{13}CO^+ J=1-0}$.

In submillimeter continuum map we can find the counterpart
of G0.11$-$0.11, although it is less prominent 
than major submillimeter features 
\citep{pierce-price2000}.
Using the same conversion from gas column density to 
submillimeter brightness as \citet{pierce-price2000},
$N(\mathrm{H}_2)= 6 \times 10^{23} \mathrm{cm^{-2}}$ corresponds to
20 Jy beam$^{-1}$ at 450 $\mu$m with 8\arcsec~ beam 
and 8 Jy beam$^{-1}$ at 850 $\mu$m with 15\arcsec~ beam.
The maps with SCUBA (Submillimeter Common-User Bolometric Array; 
Pierce-Price et al. 2000) 
show about 15--20 Jy beam$^{-1}$ at 450 $\mu$m 
and 3--4 Jy beam$^{-1}$ at 850 $\mu$m.
The estimated values are consistent 
because both estimations are based on assumptions with some uncertainty.
The gas-to-dust mass ratio may be reduced in the cloud 
because strong thermal SiO line of the cloud suggests dust evaporation.
The molecular abundance of H$^{13}$CO$^+$ may be smaller
than the value we assumed.
Moreover, inhomogeneity in the cloud may affect the conversion factors
from the observable values to the true mass of the cloud. 

The estimated column density of G0.11$-$0.11
is one of the largest ever observed even in the Galactic center region.
For several X-ray sources in the Galactic center region,
total hydrogen column densities were estimated
to be $N_{\mathrm{H}}~\lesssim~(1 - 3)\times10^{23} \mathrm{cm^{-2}}$
\citep{sakano1999,sakano2000}.
The cloud G0.11$-$0.11 shows a column density larger by at least factor of 5
than the ordinary environments in the Galactic center region.
It is extraordinarily large, being comparable only to
Sgr~B2, which is the most massive cloud in the Galaxy.

Other than G0.11$-$0.11, 
many dark features in the \textit{MSX} map are found
in the Galactic center region. 
They are called the \textit{MSX} dark clouds \citep{egan1998}.
Some of them have also been observed in the $\mathrm{H_2CO}$ line \citep{carey1998}.
Typical column densities for \textit{MSX} dark clouds 
are estimated to be $N(\mathrm{H}_2)= 10^{23-25} \mathrm{cm^{-2}}$.
Our estimated column density of G0.11$-$0.11 is as large as
the typical \textit{MSX} dark clouds; n.b., \citep{carey1998} did not observe
G0.11$-$0.11.

Finally we estimated 
the molecular mass of G0.11$-$0.11 to be $6.3\times10^5 M_{\sun}$ 
from the integrated intensity in the H$^{13}$CO$^+$ $J=1-0$ line
for $T_\mathrm{ex}=70~\mathrm{K}$.
This is consistent with the previous estimate, based on the CS $J=1-0$ line
($3.6\times10^5~M_{\sun}$, Tsuboi et al. 1997).
We should note that the previous estimate
was made on the assumption that
the CS $J=1-0$ line is moderately opaque ($\tau~\sim~1$)
and should therefore have a large uncertainty.

\subsection{The SiO-emitting Clump and its Structure}
The intensity ratio of the two SiO lines, $R_{\mathrm{SiO(2-1)/SiO(1-0)}}$
is about 0.9--1.0 
for all the features in G0.11$-$0.11 (\S~\ref{sec:ratio}). 
This value implies two possibilities;
one is that the both SiO lines are optically thick, and
the other is that both lines are optically thin and
the density of molecular hydrogen in the SiO-emitting gas
is $10^{3.7-3.8} \mathrm{cm}^{-3}$.
However, the latter case is unlikely for the following two reasons.
If the hydrogen density were $10^{3.7-3.8} \mathrm{cm}^{-3}$,
the CS lines would be optically thin.
But our estimation (\S~\ref{sec:nh}), as well as the previous
estimate \citep{tsuboi1997}, shows that the CS line is 
(at least moderately) optically thick.
Moreover, uniformity of the SiO line intensity ratio over G0.11$-$0.11
under the optically thin case
requires uniformity of the gas density 
over a scale of several parsecs in a cloud 
that, on the contrary, is known to have a complicated morphology.
Therefore, we conclude the former case to be likely:
both SiO lines are optically thick in G0.11$-$0.11.

Because the observed antenna temperature at the SiO line is much lower
than the expected gas kinetic temperature,
the beam filling factor must be smaller than unity;
i.e. the telescope beam is not filled by the emitting surface. 
Hence, we should employ a ``clumpy model'' \citep{snell1984}
to consider the physical state of G0.11$-$0.11.
Using the clumpy model,
the observed line intensity ratio depends on 
the opacity of the emitting clumps,
and the observed antenna temperature is reduced
by the beam filling factor.

\begin{figure}
\plotone{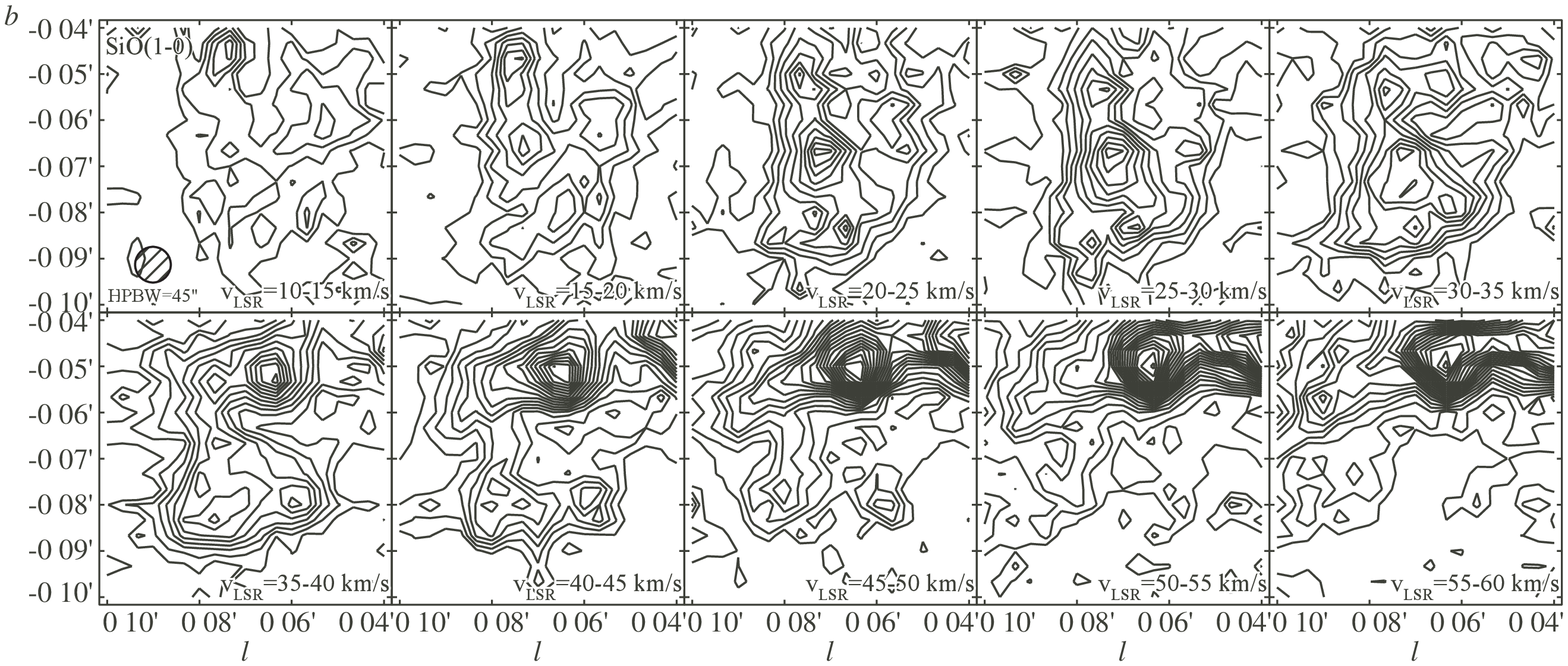}
\caption{
Same as Fig.~\ref{FigH13CO10channelMap}
but for SiO $J=1-0$
}\label{FigSiOchannelMap}
\end{figure}

\begin{figure}
\plotone{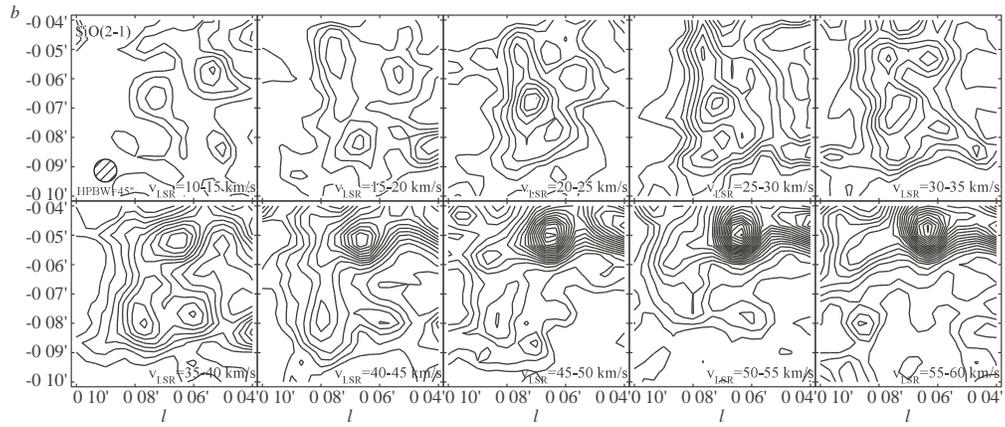}
\caption{
Same as Fig.~\ref{FigH13CO10channelMap}
but for SiO $J=2-1$
}\label{FigSiO21channelMap}
\end{figure}

\begin{figure}
\plotone{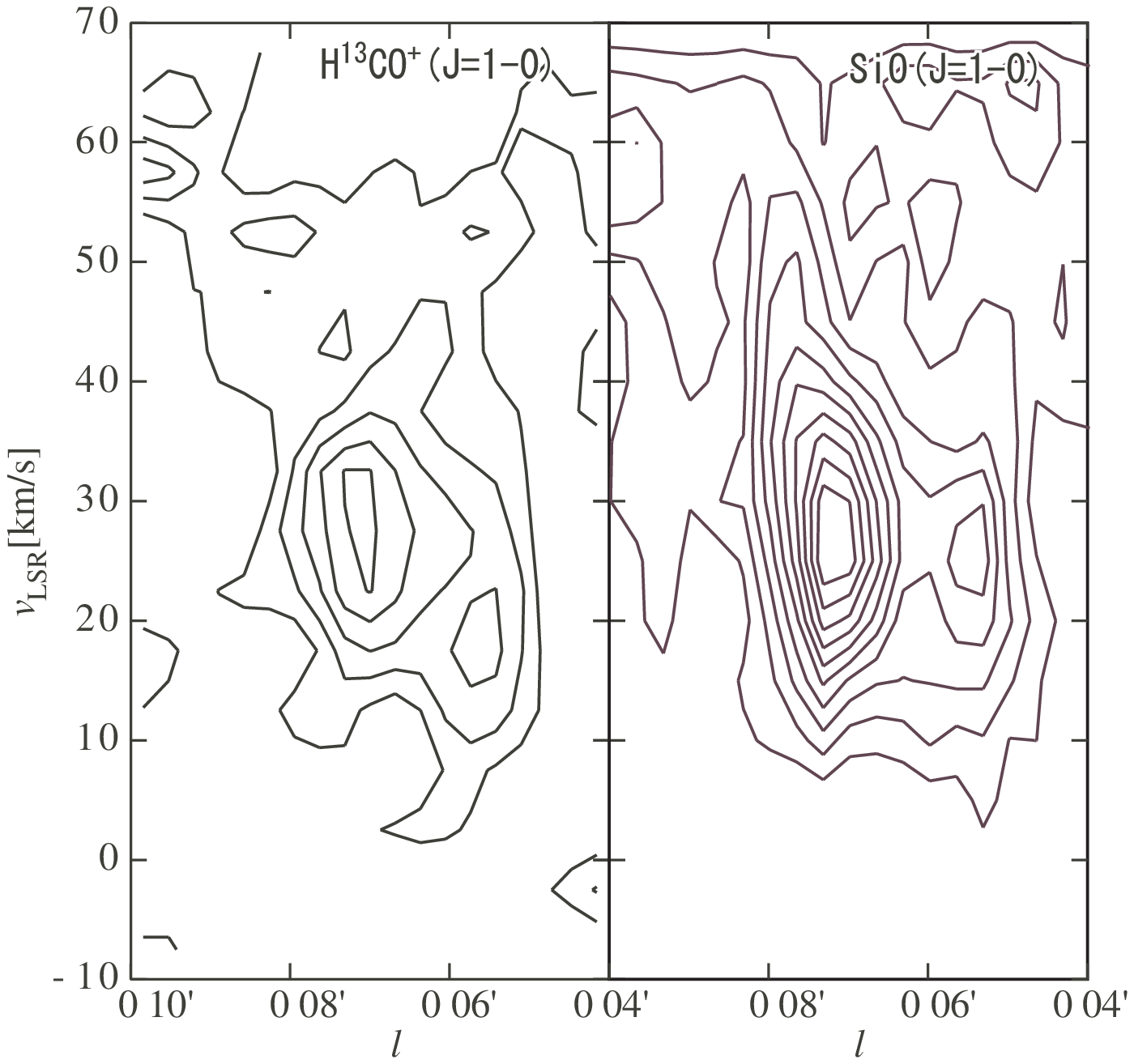}
\caption{
Position-velocity diagram 
along Galactic longitude
at $l=0\degr7\arcmin40\arcsec$
in H$^{13}$CO$^+$ $J=1-0$ (\textit{left}) and 
SiO $J=1-0$ (\textit{right}).
The angular resolution is smoothed to 45\arcsec~
but integrated over $-0\degr6\arcmin40\arcsec ~\le ~b ~\le 
~-0\degr6\arcmin20\arcsec $.
Both the first contour level and the contour interval are 
0.1 K in $T_{\mathrm{MB}}$.
}\label{FigH13CO10_lv}
\end{figure}

Using the clumpy model and an optically thick line,
we can roughly estimate some clump parameters.
In Figure {\ref{FigSiOchannelMap}}, 
we find that the main-beam brightness temperatures of
most features in G0.11$-$0.11 are $T_{\mathrm{MB}}=1-2 \mathrm{K}$.
When $T_{\mathrm{K}}=70\mathrm{K}$, the typical beam filling factor is 0.02.
Because G0.11$-$0.11 does not show a discrete clump with 35\arcsec~ beam,
there must be 10 or more clumps in a beam.
Thus, the clump diameter is smaller than 1.5\arcsec, or 0.06 pc.
The averaged gas density in a clump is then estimated to be
higher than $2\times10^8 \mathrm{cm}^{-3}$.

In this case, a clump may be unstable 
because the free-fall time of the clump 
is much shorter than the sound crossing time.
Howerver, it can be stable when the size of the clump is much smaller.
Given a beam averaged column density and beam filling factor,
the free-fall time is propotional to square root of the clump size.
On the other hand, given a gas temperature, 
the sound crossing time is propotional to the size.
Therefore, the clump can be stable when the size is smaller 
than $4\times10^{-4}$ pc 
for our estimated values.
Note that the critical size may be much larger 
if the clump is supported by magnetic field.

The clumpy model can also explain why the intensity ratios
of the optically thin (H$^{13}$CO$^+$ $J=1-0$) to thick (the CS
and SiO) lines are not significantly different 
in the southern half of the cloud.
In the clumpy model,
the shape of a line profile observed with a finite beam size is
given only by the distribution of the emitting clumps in velocity space.
In the case of a small beam filling factor,
each emitting clump does not screen other clumps
even if the line is optically thick.
Therefore, optically thick lines show almost the same profile shape
as optically thin lines.

The CS line intensity observed by \citet{tsuboi1997} shows that
the main-beam brightness temperature in the CS $J=1-0$ line is
brighter than that in the SiO lines by a factor of 3,
although both the lines are optically thick in G0.11$-$0.11. 
This means that the beam filling factor at the SiO line must be smaller than 
that at the CS line. However, in the case of the simplest clumpy model,
in which every emitting clump is assumed to be 
isothermal and uniform in density, 
the line intensity ratio of SiO to CS is expected to be unity
because the excitation parameters of SiO and CS are very close.
To resolve this discrepancy, 
we deduce that
the emitting gas clump has a steep abundance gradient in SiO
and that the typical size of an optical thick surface 
(or beam filling factor) that emits the SiO line is much smaller than
the corresponding emission surface of the CS line.
With the same clump temperature and density,
the optical depth in the SiO lines can vary, depending on the SiO abundance.
In fact, the observed SiO abundance, $X\mathrm{(SiO)}$, differs 
by several orders of magnitude for molecular gas in the Galactic
disk region: e.g., $X(\mathrm{SiO})\simeq10^{-12}$ for quiescent dark clouds 
\citep{ziurys1989} and $X(\mathrm{SiO})=10^{-7}-10^{-8}$ in bipolar
outflows of star-forming regions
\citep{martinpintado1992,schilke1997,gueth1998}.

The ratio of the projected area of the SiO-emitting part 
to that of the CS-emitting part is the same as the value of the intensity
ratio of these lines 
because both lines are optically thick.
We find that the intensity ratio of SiO $J=1-0$ to CS $J=1-0$ is uniform 
over the cloud.
This uniformity implies that
the ratio of the projected areas in a clump is uniform over the cloud.
Thus, the abundance gradient of SiO in the emitting clump is presumably
due to a mechanism on a scale much larger than the whole cloud.
Consequently, clump internal structure should be the same
over the whole cloud.
In this case, 
it is a reasonable assumption that 
the emitting clump is spherically symmetric.

In the case of a spherically symmetric clump,
the SiO-emitting part should be at smaller radius than the CS-emitting part.
It follows that the emitting clump is hotter in the innermost part 
because the SiO abundance is believed to increase 
in hot (e.g. shock heated) gas.
Even in this case, our discussion is valid, 
although our model is inconsistent to the simplest clumpy model;
the main-beam brightness temperature of an optically thick clump is the same
when the product of the surface area and the actual brightness temperature
is the same.

The large opacity even in dense gas tracers and large extinction 
even in mid-infrared suggest that the cooling time may be longer
than in cores in star-forming regions in the Galactic disk region.
Using virial mass analysis, 
\citet{sawada2001} show that
molecular clouds in the Galactic center region
are under a high pressure of $n T_\mathrm{K} = 10^5 \mathrm{K~cm^{-3}}$. 
Emitting clumps in G0.11$-$0.11 
may be compressed by this external pressure.

From our observations, G0.11$-$0.11 is found to be likely composed of
many hot and dense clumps, which can hardly be
cooled down because of large extinction even in infrared.
This condition is greatly different from that of star-forming clouds
in the Galactic disk region.
Under such condition, 
star formation should be very different.
This may be a reason why a dense cluster of massive stars is seen only
in the Galactic center region.
Although G0.11$-$0.11 is a good site for investigating this speculation,
high-resolution observations in rarer molecules 
such as CS isotopes are required to unveil optically thick clumps.


\acknowledgments

The authors express their thanks to I. M. Stewart for his linguistic help.
The authors thank to the referee for suggestions that
improved the paper.



Facilities: \facility{No: 45 m}.




\clearpage



\clearpage

\clearpage

\clearpage

\clearpage


\end{document}